\documentclass[%
 reprint,
 amsmath,amssymb,
 aps,
]{revtex4-2}

\usepackage{graphicx}
\usepackage{dcolumn}
\usepackage{bm}
\usepackage{xcolor}
\usepackage{siunitx}

\begin{document}    

\preprint{APS/123-QED}

\title{Sound propagation in a Bose-Fermi mixture: from weak to strong interactions}

\author{Krutik Patel, Geyue Cai, Henry Ando, and Cheng Chin}
\affiliation{The James Franck Institute, Enrico Fermi Institute, and Department of Physics, \\ The University of Chicago, Chicago, IL 60637, USA}

\begin{abstract} 
Particle-like excitations, or quasi-particles, emerging from interacting fermionic and bosonic quantum fields underlie many intriguing quantum phenomena in high energy and condensed matter systems. Computation of the properties of these excitations is frequently intractable in the strong interaction regime.  Quantum degenerate Bose-Fermi mixtures offer promising prospects to elucidate the physics of such quasi-particles. In this work, we investigate phonon propagation in an atomic Bose-Einstein condensate immersed in a degenerate Fermi gas with interspecies scattering length $a_\text{BF}$ tuned by a Feshbach resonance. We observe sound mode softening with moderate attractive interactions. For even greater attraction, surprisingly, stable sound propagation re-emerges and persists across the resonance. The stability of phonons with resonant interactions opens up opportunities to investigate novel Bose-Fermi liquids and fermionic pairing in the strong interaction regime.
\end{abstract}

\maketitle

Interactions between excitations of bosonic and fermionic quantum fields play an important role in understanding fundamental processes in high energy and condensed matter physics. In quantum electrodynamics, for example, the coupling between the photon and virtual electron-positron pairs polarizes the vacuum, which  contributes to Lamb shifts \cite{Lamb1947} and the anomalous magnetic moments  of the electron and the muon \cite{Schwinger1948}. In condensed matter, interactions between phonons and electrons are central to Cooper pairing in conventional superconductors \cite{Bardeen1957}, as well as charge ordering and superconductivity in strongly correlated materials \cite{Giustino2017,Keimer2015}.

Ultracold mixtures of atomic Bose and Fermi gases offer a complementary experimental platform for elucidating these quantum phenomena. Cold atoms are exceptionally flexible, allowing for the control of interactions between the atomic species using Feshbach resonances \cite{Chin2010}.  These capabilities have been used to study phase transitions in  lattices \cite{Gunter2006,Ospelkaus2006,Sugawa2011}, polarons \cite{Yan2020,Fritsche2021}, and superfluid mixtures \cite{Marion2015,Roy2017}. Many exciting theoretical predictions for quantum simulation remain to be tested, e.g. Refs. \cite{Pazy2005,Efremov2002,Banerjee2012}.

\begin{figure}\center
\includegraphics[width=0.48\textwidth]{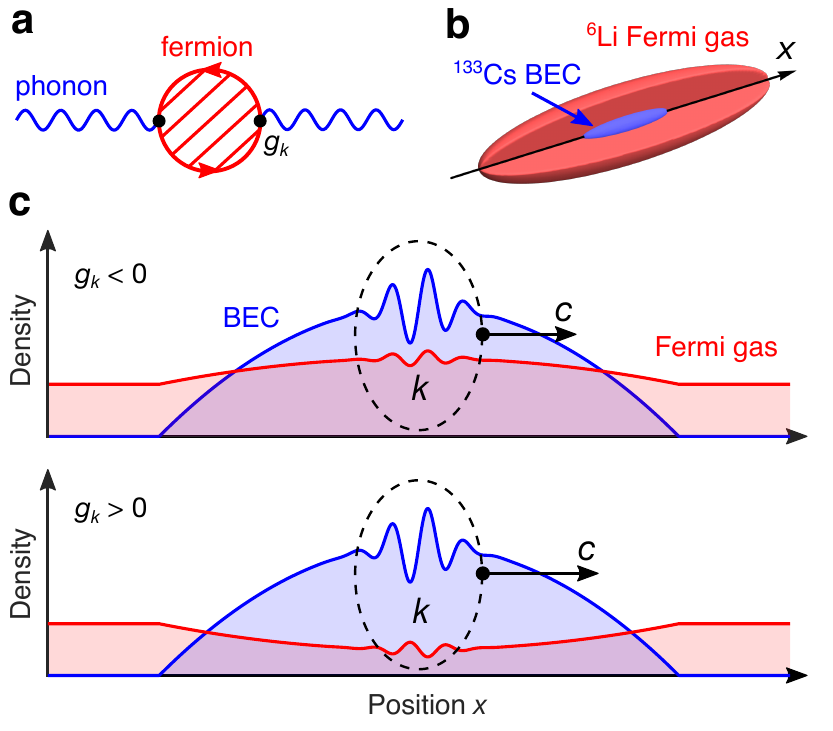}
\caption{\label{fig:soundcartoon} Bosonic quasi-particles (phonons) coupled to a fermionic quantum field. (a) Diagrammatic representation of phonons (blue) coupled to excitations of a fermionic field (red).  The lowest order diagram contains a single loop and is second order in the phonon-fermion coupling $g_k$. Higher order corrections are indicated by the hatched area. (b) In our experiment, a cigar-shaped Bose-Einstein condensate (BEC) of cesium-133 is immersed in a much larger  degenerate Fermi gas of lithium-6. (c) As a phonon with momentum $k$ (black dashed ellipse) propagates, the coupling results in the density modulation of both species and the modification of the sound speed $c$. }
\end{figure}

In this work, we investigate sound propagation in a quantum degenerate Bose-Fermi mixture from the weak to the strong interaction regime. We optically excite density waves in the gases and measure their velocities and damping rates from \textit{in situ} images of the  Bose-Einstein condensate (BEC). We see significant changes in  the speed of sound for interspecies attraction and negligible shifts for repulsion. This asymmetry indicates strong deviation from the perturbation prediction. Intriguingly, we find stable propagation of sound waves in mixtures with resonant interspecies interactions. This observation offers promising prospects to explore new quantum phases of Bose-Fermi mixtures in the strong interaction regime.

\begin{figure*}\centering
\includegraphics[trim=0in 0.2in 0in 0in,clip,width=\textwidth]{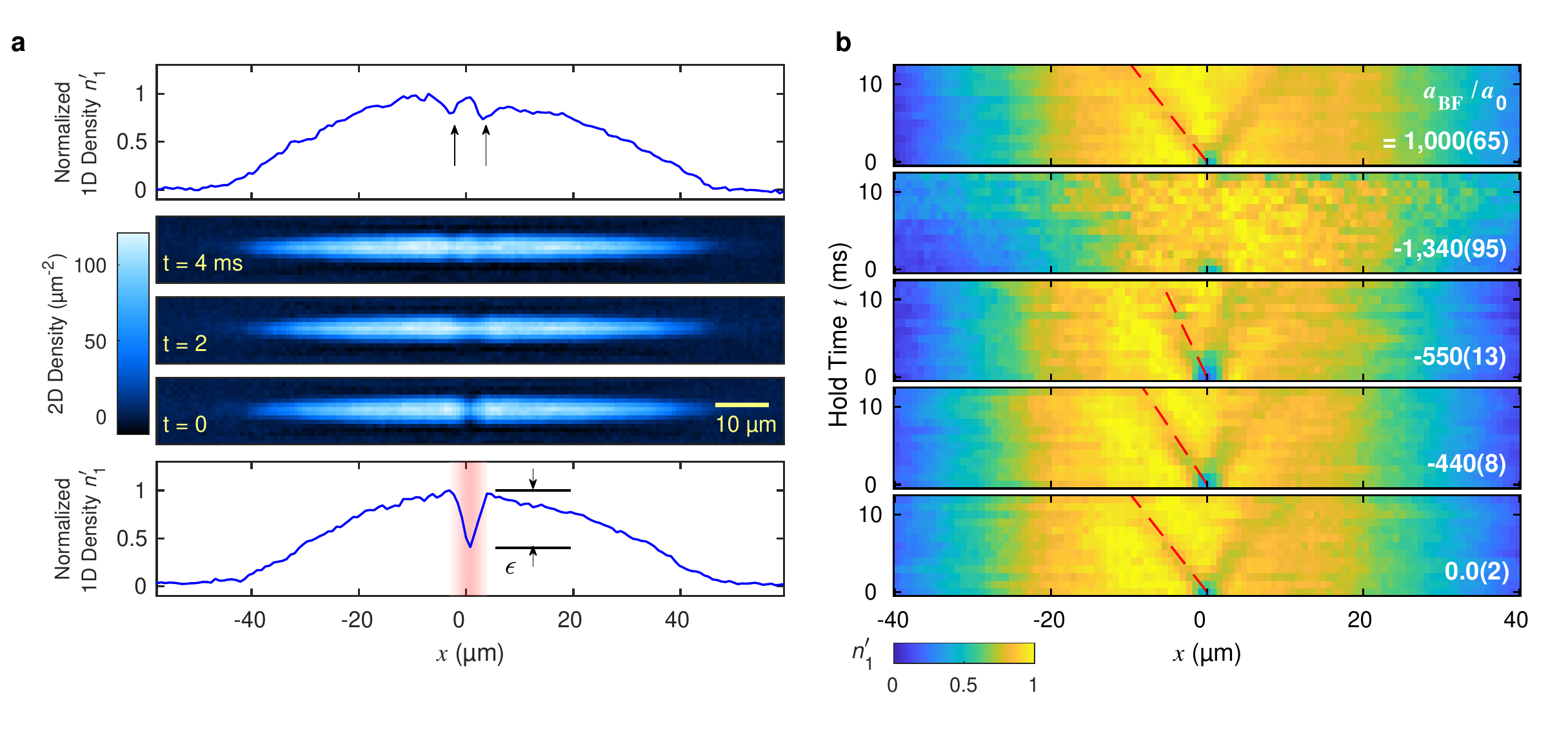}
\caption{\label{fig:sampledata} Excitation and \textit{in situ} imaging of density waves. (a) A local density depletion $\epsilon$ is created in the center of the cesium BEC by a projected laser beam (bottom panel, red shaded area). The optical potential is abruptly switched off at $t=0$ and the density dip splits into density waves propagating in opposite directions (top panel, black arrows). Average column densities are shown for three values of the hold time $t = 0, 2, 4 $ ms  along with sample normalized one-dimensional (1D) densities $n'_1$ for $t=0$ ms and $t=4$ ms. Data is shown for the Cs-Li Bose-Fermi mixture with interspecies scattering length $a_{\text{BF}}= -335\,a_0$. (b)  Normalized 1D densities $n'_1$ show density wave dynamics for mixtures prepared at various interspecies scattering lengths. Red dashed lines are guides to the eye.}
\end{figure*}

The Hamiltonian for the phonons coupled to a single-component  Fermi gas is given by \cite{Viverit2002,Enss2009}

\begin{equation}
 H =  \sum_k\epsilon_k^\text{F}c_k^{\dagger}c_k + \sum_k\hbar\omega_k\alpha_k^{\dagger}\alpha_k + \sum_{k,q}g_k(\alpha_k+\alpha^{\dagger}_{-k})c^{\dagger}_qc_{q-k},
\end{equation}

\noindent  where $\epsilon_k^\text{F}$ is the dispersion of the fermions, $\hbar$ is the reduced Planck's constant, $\omega_k$ is the phonon dispersion, $g_k$ is the phonon-fermion coupling constant, $c_k$ and $\alpha_k$ refer to fermion and phonon annihilation operators respectively, and $k$ and $q$ are momenta (see Fig.~1a). In our degenerate Bose-Fermi mixture, the kinetic energy of a bare fermion is $\epsilon_k^\text{F} = \hbar^2k^2/2m_\text{F}$, where $m_\text{F}$ is the fermion mass. The bare phonons are low energy excitations of the BEC with the Bogoliubov dispersion \cite{Pethick2002}   $\omega_k \approx c_0 k$, where the sound velocity $c_0 = \sqrt{g_\text{BB}n_\text{B}/{m_\text{B}}}$ is determined by the boson-boson coupling constant $g_\text{BB}$, condensate density $n_\text{B}$, and boson mass $m_\text{B}$. The phonon-fermion coupling constant is $g_k = g_{\text{BF}}\sqrt{n_\text{B}\hbar k^2/2m_\text{B}\omega_k}$ \cite{Enss2009,Supplement}, where $g_\text{BF} = 2\pi\hbar^2a_\text{BF}/m_{r}$ is the interspecies coupling constant, $a_\text{BF}$ is the interspecies scattering length and $m_r$ is the reduced mass of the two unlike atoms. The phonon-fermion  coupling $g_k$ can thus be tuned by controlling $a_\text{BF}$ using an interspecies Feshbach resonance (see Fig. 1c).

Perturbation theory shows that the velocity of phonons is reduced when the BEC interacts weakly with the Fermi gas. This can be understood as a result of a  fermion-mediated interaction between bosons analogous to the Ruderman-Kittel-Kasuya-Yosida mechanism \cite{Ruderman1954,De2014}. The mediated interaction has been observed in cold atom experiments \cite{OurNature,Edri2020}. To leading order in $g_{\text{BF}}$, the sound velocity is predicted to be \cite{Yip2001}

\begin{equation} \label{rkkyc}
c = c_0\sqrt{1-\frac{3}{2}\frac{g_\text{BF}^2}{g_{\text{BB}}}\frac{n_\text{F0}}{E_\text{F0}}},
\end{equation}

\noindent 
where $n_\text{F0}$ and $E_\text{F0}$ are the density and Fermi energy of the Fermi gas in the absence of the condensate.  This correction is quadratic in the coupling strength $g_\text{BF}$, and corresponds to the one-loop diagram shown in Fig 1a. The sound speed is expected to be reduced regardless of the sign of the interspecies coupling strength $g_{\text{BF}}$. The perturbation result is valid in the weak coupling regime  $|g_{\text{BF}}n_{\text{B}}|\ll E_{\text{F0}}$.

\begin{figure*}
\includegraphics[trim=0.4in 0.5in 0in 0in,clip,width=\textwidth]{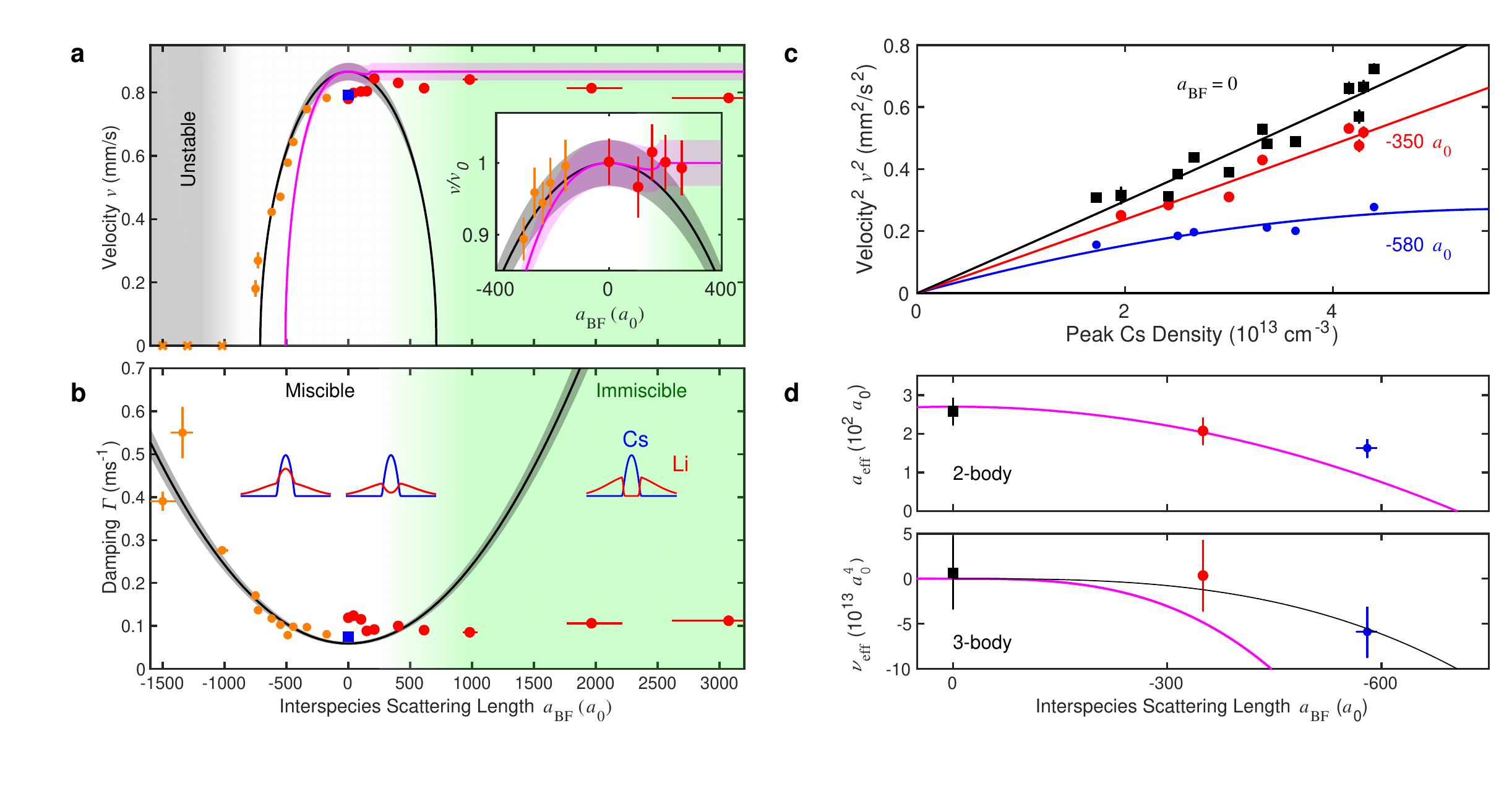}
\caption{\label{fig:cvsabf} Sound speeds and damping rates at different interspecies scattering lengths and boson densities.  (a) Orange data points indicate samples prepared on the attractive side of the Feshbach resonance $a_{\text{BF}}<0$. Red data points are prepared on the repulsive side with $a_{\text{BF}}>0$. Measurement for a bare BEC without femions is shown as the blue square. The crosses indicate samples with no stable sound propagation. Inset shows the ratio of density wave velocities for samples prepared with and without the fermions. The ratios are obtained from the separation of density waves after 8 ms hold time. Calculations from perturbation (black line) and mean-field (magenta line) theory are shown for comparison. The green shaded area represents the phase separation region. The grey area indicates the region where no stable sound propagation is observed. (b) Damping rates of the density waves are compared with the perturbative prediction (black line) evaluated for momentum  $k=2\pi/(4 \mu$m) \cite{Supplement}. Insets: Cartoon representation of the Cs (blue) and Li (red) density profiles in different regimes. (c) Density wave velocities for BECs prepared without the Fermi gas (black squares) and with the Fermi gas at $a_\text{BF}=-350\,a_0$ (red circles) and  $a_\text{BF}=-580\,a_0$ (blue circles). Lines are fits of the data to a model with both two- and three-body effective interactions between bosons (see text). (d) Colored circles are the effective scattering lengths and hypervolumes extracted from panel (c). The magenta lines are the mean-field predictions and the black line is a cubic fit  to the data.  The vertical error bars on the data in (a)-(c) are standard errors calculated from  fits to averaged experimental density profiles. The horizontal error bars on the data in panels (a), (b), and (d) represent the 1-$\sigma$ uncertainty of the scattering length \cite{Supplement}. The shaded regions around the theory calculations in panels (a) and (b) indicate the ranges of the predictions \cite{Supplement}. The error bars in panel (d) are standard errors calculated from fits to the data in panel (c).}
\end{figure*}

At stronger interactions, the density profile of each species can be significantly modified by the other species. This effect can be captured in a mean-field model. Under the Thomas-Fermi approximation for both species, the local mean-field chemical potential of the bosons depends on the fermion density as \cite{Supplement}

\begin{equation} \label{mfmodel}
 \mu_{\text{TF}} = g_\text{BB}n_\text{B} + g_\text{BF}n_{\text{F0}}\left(1 - \frac{g_\text{BF}n_\text{B}}{E_\text{F0}} \right)^{3/2},
 \end{equation}
\noindent

\noindent where the second term is  set to zero when the mean-field interaction energy exceeds the Fermi energy, $g_{\text{BF}}n_\text{B} > E_\text{F0}$.  In our system, it is a good approximation that the light fermions (Li) follow the heavy bosons (Cs) adiabatically.  This permits the evaluation of the mean-field sound speed $ c=\sqrt{n_\text{B}/m_\text{B}\kappa}$
in terms of the effective compressibillity $\kappa = \partial n_\text{B}/ \partial{\mu_{\text{TF}}}$ as

\begin{equation}
\label{mfeq1}
 c = c_0\sqrt{1 - \frac{3}{2}\frac{g_\text{BF}^2}{g_\text{BB}}\frac{n_\text{F0}}{E_\text{F0}}\sqrt{1 - \frac{g_\text{BF}n_\text{B}}{E_\text{F0}}}}.
\end{equation}

\noindent Compared to Eq.~(2), the additional factor in Eq.~(4) captures the density changes in the mixture caused by interspecies interactions.

Our experiments begin with mixtures of a pure BEC of 30,000 $^{133}$Cs atoms and a degenerate Fermi gas of 8,000 $^6$Li atoms. Both species are spin polarized into their lowest  hyperfine ground states \cite{Supplement,OurPRL}. For Cs, this state is adiabatically connected to $|F=3,m_F=3\rangle$ and for Li it is connected to $|F=1/2, m_F=1/2\rangle$  at low magnetic fields, where $F$ is the total angular momentum quantum number and $m_F$ is the magnetic quantum number. The mixture is trapped in a single beam optical dipole trap at wavelength $1064$ nm  with trap frequencies $\omega_{\text{Cs}} = 2\pi \times (6.53,100,140)$ Hz and $\omega_{\text{Li}} =  2\pi \times (36,330,330)$ Hz in the axial and two transverse directions.  The bosons and fermions have a temperature of about 30 nK and chemical potentials of about $k_\text{B} \times 30$ nK  and $k_\text{B} \times  300$ nK respectively, where $k_\text{B}$ is the Boltzmann constant. In the dipole trap, the  BEC is fully immersed in the degenerate Fermi gas (see Fig. 1b). We tune the interspecies scattering length near a narrow Feshbach resonance at magnetic field 892.65 G \cite{Supplement,Tung2013,Johansen2017}. Across the resonance, the boson-boson interactions are moderately repulsive with a nearly constant scattering length $a_\text{BB} = 270~a_0$ \cite{Berninger2013}, where $a_0$ is the Bohr radius. At these temperatures, the interactions between the single component Li atoms are negligible. In our experiment, the mixture is prepared in the weak coupling regime, where the interspecies scattering length is $|a_{\text{BF}}| < 200 a_0$.

\begin{figure*}\center
\includegraphics[width=\textwidth]{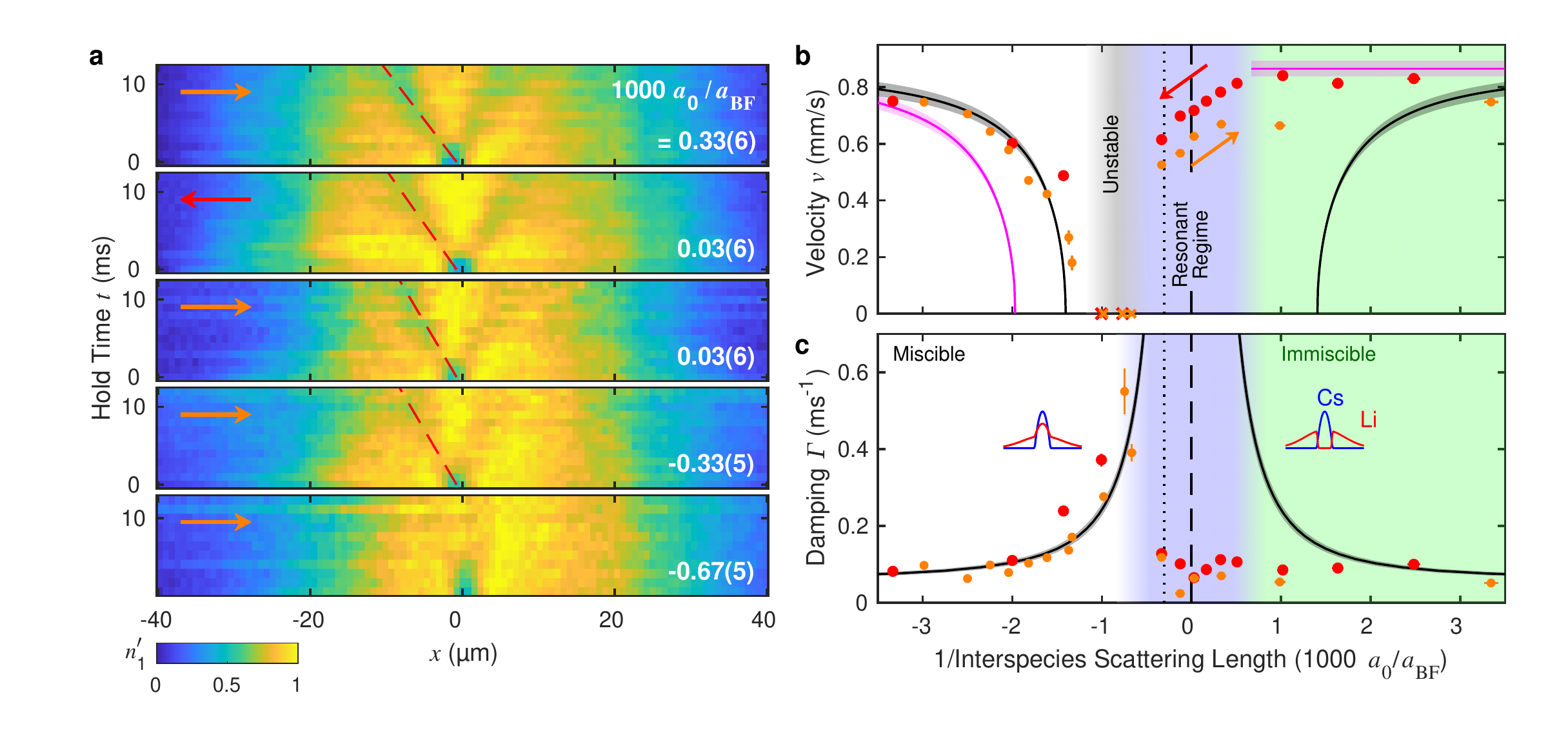}
\caption{\label{fig:cres} Sound propagation across the Feshbach resonance. (a) Normalized 1D densities illustrating the revival of sound propagation at strong interactions based on the same experimental procedure as in Figs.~2 and 3. The arrows on each data set indicate whether the system is ramped towards the resonance starting from the attractive side (orange arrow) or repulsive side (red arrow). Red dashed lines are guides to the eye. (b) Density wave velocity of the Li-Cs mixture across the Feshbach resonance. Data taken from samples prepared on the attractive (repulsive) side are orange (red) in color. The arrows indicate the direction of the scattering length ramp. The blue and green regions indicate the resonant  and phase separation regimes respectively.  (c) Damping from the same data set. The black and magenta lines are the same perturbation and mean field predictions as shown in  Fig.~3. The vertical error bars in panels (b) and (c) are standard errors from fits to averaged density profiles. 
The shaded regions around the theory curves in panels (b) and (c) indicate the ranges of the predictions \cite{Supplement}. The vertical dashed line shows the position of the Feshbach resonance. The vertical dotted line shows the position of the Efimov resonance reported in Ref.~\cite{Johansen2017}.}
\end{figure*}

To study sound propagation in our system, we optically excite density waves in the mixture \cite{Meppelink2009,Andrews1997,Joseph2007}.
We introduce a narrow repulsive potential barrier of width $\delta=4$~\SI{}{\micro\meter} by projecting blue-detuned light onto the center of the BEC, resulting in a density dip. 
We then switch the magnetic field to the target scattering length. After 5 ms, when the magnetic field stabilizes, we turn off the optical barrier and record the dynamics of the density waves after various hold times $t$ \cite{Supplement}. We observe that the initial density depletion splits into two density waves that counter-propagate at the same speed along the axial direction (see Fig.~\ref{fig:sampledata}a). From the images, we extract the velocity $v$ and damping rate $\Gamma$ of the density waves \cite{Supplement}. We repeat the experiment at different interspecies interaction strengths (see Fig.~\ref{fig:sampledata}b).

The density wave velocity $v$ in a bare elongated condensate is given by the sound speed $c_0$ through \cite{Kavoulakis1998,Supplement}

 \begin{equation}
v \approx \frac{c_0}{\sqrt{2}}\sqrt{1-\frac{\epsilon}{2}},
\label{depletionc}\end{equation}

\noindent where $\epsilon$ is the initial density depletion due to the potential barrier (see Fig. 2a) and $c_0$ is the sound speed at the center of the BEC.

In the presence of fermions, we measure the dependence of the density wave velocity on the initial density depletion $\epsilon$ and find agreement with Eq.~(5) \cite{Supplement}. Thus, we adopt Eq.~(5) to link the density wave velocity to the sound speed.  In the following experiments, the initial density depletion is set to $\epsilon=0.5$.

 We summarize the measured density wave velocities and damping rates in Figs. 3a and 3b. As we increase the interspecies attraction from zero, the density waves propagate slower and decay faster. The enhanced damping of the density waves is consistent with the perturbation calculation for a zero-temperature Bose-Fermi mixture \cite{Viverit2002, Supplement}. When the scattering length exceeds the critical value of $a_{\text{c}}=-790(10)\,a_0$ \cite{Supplement}, we no longer observe stable propagation of sound.  
 Our finding is consistent with the sound mode softening in the Bose-Fermi mixture with increasing attraction. Our measured critical value shows clear deviations from the perturbation prediction $-710\,a_0$ and the mean field prediction $-510\,a_0$ for the collapse of the mixture \cite{Molmer1998}.

For repulsive interspecies interactions, on the other hand, the density waves propagate with low damping and no significant change in velocity over the range we explore (see Figs.~3a and 3b). This is in stark contrast to our observations for attraction. The clear asymmetry with respect to the sign of the interaction goes beyond the perturbation prediction, see Eq.~(2), which only depends on the square of the scattering length $a_{\text{BF}}^2$.

The asymmetry can be understood from the mean-field picture. For attractive interactions, fermions are pulled into the BEC, and the higher fermion density further reduces the sound velocity.  On the other hand, for repulsion, fermions are expelled from the BEC, reducing their effect on the sound propagation. For strong enough repulsion, the bosons and fermions are expected to phase separate \cite{Molmer1998,Viverit2000,Lous2018}. The observed nearly constant sound velocity for strong repulsion is consistent with the picture that most fermions are expelled from the condensate. For our system, the mean field model predicts phase separation near the scattering length $a_{\text{BF}}\approx 180\, a_0$. 

This asymmetry comes fundamentally from effective few-body interactions in the BEC that go beyond the perturbation calculation \cite{Belemuk2007, Enss2020}. The change of the density overlap, described in the mean-field picture, is a consequence of the few-body interactions. The effective boson-boson-boson three-body interaction strength can be experimentally characterized by writing the chemical potential in orders of the boson density

\begin{equation}
\mu = g_2n_\text{B} + g_3n_\text{B}^2 + ... ,
\label{gcoeff}\end{equation}
 where  $g_2 = 4\pi\hbar^2a_\text{eff}/m_\text{B}$ and $g_3 \equiv \hbar^2\nu_\text{eff}/m_\text{B}$ are effective two- and three-body coupling constants between bosons, $a_\text{eff}$ is the effective scattering length, and  $\nu_\text{eff}$ is the effective scattering hypervolume. From the effective chemical potential $\mu$  we obtain the sound speed as $c \approx \sqrt{(g_2n_\text{B} + 2g_3n_\text{B}^2)/m_\text{B}}$.

To determine the effective two- and three-body interaction strengths, we measure the density wave velocity at various boson densities and scattering lengths. The results are shown in Fig.~\ref{fig:cvsabf}c. From fits to the density wave velocities and Eqs.~(5) and (6), we extract the effective scattering length $a_{\text{eff}}$ and effective scattering hypervolume $\nu_{\text{eff}}$ (see Fig.~3d).

 As the interspecies attraction increases, we observe a reduction of the effective scattering length, consistent with Ref.~\cite{OurNature}, and an emerging scattering hypervolume. Mean-field theory predicts $\nu_\text{eff} = \lambda a_\text{BF}^3$ with  $\lambda \approx 159 k_\text{F}^{-1}$ set by the Fermi momentum and mass ratio ~\cite{Supplement}. Fitting the data, we determine $\lambda = 35(8) k_\text{F}^{-1}$, see Fig.~3d. This value shows clear deviation from the mean field prediction. Notably, the three-body interaction $g_3n_\text{B}^2 \propto a_{\text{BF}}^3$ is the leading order process that breaks the symmetry between positive and negative scattering length.

By ramping our magnetic field across the Feshbach resonance, we explore the sound propagation in the strong interaction regime, where the scattering length exceeds all length scales in the system. Surprisingly, we observe stable sound propagation with low damping for all scattering lengths $|a_{\text{BF}}|>3{\small,}000\,a_0$ (see Fig.~4) regardless of which side of the resonance the samples are initially prepared on \cite{Supplement}. We label this range the resonant regime.  Examples of the sound propagation in the resonant regime are shown in Fig.~\ref{fig:cres}a. An interesting scenario occurs when we approach the resonance from the attractive side. The damping rate increases as the sound velocity approaches zero for stronger attraction until the sound propagation becomes unstable at the critical value $a_\text{c} = -790(10)\,a_0$. Then, between $a_c$ and $a_\text{BF} = -3{\small,}000\,a_0$ the system does not exhibit stable sound propagation. For even stronger attraction $a_\text{BF} < -3{\small,}000\,a_0$, intriguingly, stable sound propagation re-emerges and persists across the Feshbach resonance with a  damping and sound velocity comparable to weakly interacting samples. Additional data over this region is presented in the supplement \cite{Supplement}.

The stable sound propagation we observe across the interspecies Feshbach resonance goes beyond the mean-field picture and offers promising prospects for future discoveries in the strong-coupling regime. The re-emergence of the sound propagation occurs near the Efimov resonance at the scattering length $a_{\text{BF}} = -3,330\,a_0$  \cite{Johansen2017}. Theoretically an Efimov resonance can induce an effective two-body repulsion \cite{Enss2020} and stabilize sound propagation. Also, at strong interactions, mean-field corrections are predicted to support a novel quantum droplet phase for scattering lengths $a_\text{BF}<-750\,a_0$ \cite{Rakshit2019}. Finally, at strong coupling, $p$-wave fermionic superfluidity is conjectured when fermions are paired through the exchange of bosonic excitations \cite{Enss2009,Efremov2002,Kinnunen2018}, which we estimate would occur in our system in the range $a_{\text{BF}} = -2,000\,a_0$ to $-10,000\,a_0$. The stable phonon propagation we observe near the Feshbach resonance offers promising prospects to explore these intriguing physics with strongly interacting Bose-Fermi mixtures. For example, experimentally probing the interspecies correlations can help elucidate the mechanism that stabilizes the sound mode in the resonant regime.

We thank B. Evrard, J. Ho, and E. Mueller for valuable discussions. We thank X. Song and C. Li for assistance with the numerical simulations.  We thank N. C. Chiu for technical support. We thank M. Rautenberg for valuable discussions and assistance with the experiment. This work was supported by the National Science Foundation under Grant No. PHY-1511696 and PHY-2103542, by the Air Force Office of Scientific Research under award number FA9550-21-1-0447, and by the National Science Foundation Graduate Research Fellowship under Grant No. DGE 1746045.

\bibliography{apssamp.bib}
\bibliographystyle{apsrev4-1}

\clearpage
 \renewcommand\thefigure{S\arabic{figure}}

\setcounter{figure}{0}   

\maketitle

\onecolumngrid
\begin{center}

\begin{large}
\textbf{Supplementary Material for\\Observation of sound propagation in a strongly interacting Bose-Fermi mixture}
\end{large}\\
Krutik Patel, Geyue Cai, Henry Ando, and Cheng Chin\\
\textit{ The James Franck Institute, Enrico Fermi Institute and Department of Physics,\\ The University of Chicago, Chicago, IL 60637, USA}
\vspace{1cm}
\twocolumngrid
\end{center}

\subsection*{A.\,Experimental set-up and procedures}

We perform the experiments with both Cs and Li atoms prepared in their lowest hyperfine ground state. Cs atoms are initially polarized in the $|F, m_F \rangle = |3, 3\rangle$ state at a low magnetic field and Li atoms are polarized into the $|1/2, 1/2\rangle$ state, where $F$ is the total angular momentum and $m_F$ is the magnetic quantum number. We then adiabatically ramp the magnetic field near the Feshbach resonance at 892.65 G, and both species remain in their lowest internal state. A more detailed discussion of the system preparation can be found in Ref. \cite{OurPRL}. From our measurements of trap frequencies and beam parameters, we estimate a possible displacement between the vertical centers of each cloud of about 8 microns due to gravity. However, the mean-field potential felt by the Li due to the Cs has a trapping effect on the attractive side of resonance that improves the overlap of the two species. 

In Fig.~\ref{fig:avsb} we show the Cs-Cs and the Li-Cs scattering length as a function of magnetic field in the range where we perform the experiments.  The models for the scattering length are from Refs.~\cite{Berninger2013,OurPRA,Ulmanis2015} and have been adjusted based on experimental measurements \cite{OurNature,Johansen2017}.

\begin{figure}\center
\includegraphics[trim=0in 0.2in 0.1in 0.2in,clip,width=0.48\textwidth]{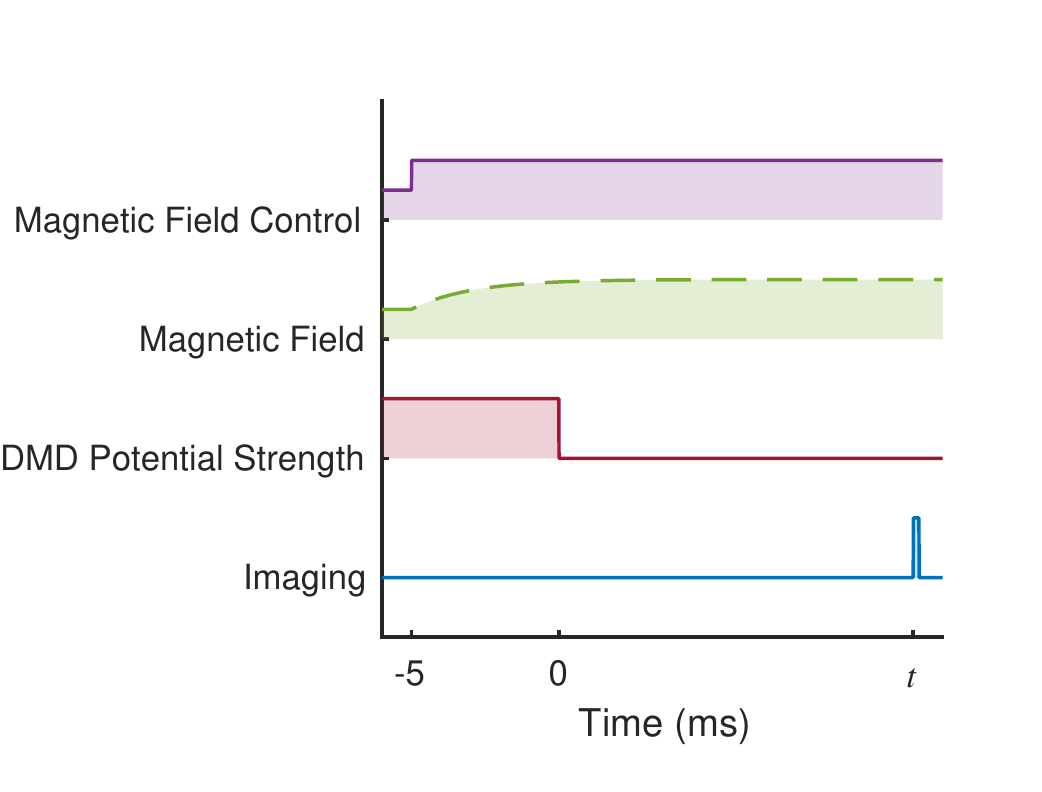}

\caption{\label{fig:timing} The experimental timing sequence for exciting and imaging density waves. We switch the magnetic field control (purple line) to the target value at $-5$ ms. The magnetic field (dashed green line) settles within 0.3\% of its final value at 0 ms. At this time we turn off the DMD potential (red line), which initiates density depletion dynamics. We image the sample after a hold time $t$ (blue line), which ranges from $0$ to $12$ ms. This timing sequence is employed for all experiments in Figs. 2, 3, and 4.}
\end{figure}

\begin{figure}\center
\includegraphics[trim=2.2in 3.9in 2.2in 3.9in,clip,width=0.48\textwidth]{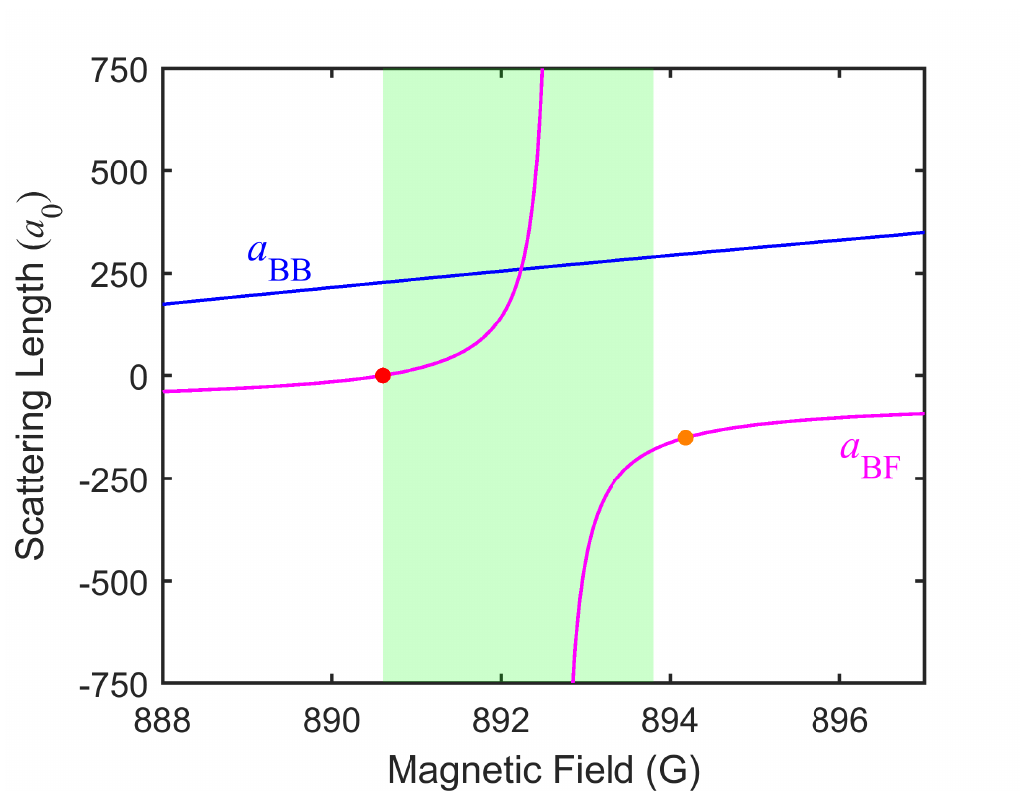}

\caption{\label{fig:avsb} Interaction strength between atoms. The scattering length between the Cs atoms $a_\text{BB}$ is shown in blue and the Li-Cs scattering length $a_\text{BF}$ is shown in magenta near the interspecies Feshbach resonance. The Cs-Cs scattering length is from the model in Ref. \cite{Berninger2013} adjusted by measurements made in Ref. \cite{OurNature}. The Li-Cs scattering length is from the model of Ref. \cite{OurPRA} adjusted by measurements in Ref. \cite{Johansen2017}. The red and orange circles indicate the initial magnetic field before the ramp to each target interspecies scattering length for data prepared on the repulsive and attractive side of the resonance, respectively. The shaded green area indicates the region of scattering lengths probed in this work.}
\end{figure}

To perform an experiment at a target interspecies scattering length $a_\text{BF}$, we first prepare the mixture at either $a_{\text{BF}} = -180\,a_0$ on the attractive side or $a_{\text{BF}} = 120\,a_0$ on repulsive side of resonance \cite{OurPRL}. Then, we ramp the magnetic field to the target value in two steps. For samples initially prepared on the attractive side, shown as orange circles in Figs. \ref{fig:cvsabf}, \ref{fig:cres}, \ref{fig:depletion_abf}, and \ref{fig:loss}, we first ramp to $a_{\text{BF}} = -150\, a_0$ in 110 ms (see Fig.~\ref{fig:avsb}) then hold for 15 ms. We then switch the magnetic field to the target value and allow the magnetic field 5 ms to settle before we turn off our optical barrier. A timing diagram of the final portion of this sequence is given in Fig. \ref{fig:timing}.For samples prepared on the repulsive side (red circles in Figs. \ref{fig:cvsabf}, \ref{fig:cres}, \ref{fig:depletion_abf}, and \ref{fig:loss}), we ramp first to $a_{\text{BF}} = 0\,a_0$, then switch to the target value, following the same timing procedure. For each data point, we determine the magnetic field using microwave spectroscopy of the $|3,3\rangle \leftrightarrow | 4,4 \rangle$ transition in the ground state manifold of Cs.

The field settling time of 5 ms is fast enough that the Cs  density profile remains approximately unchanged (see Sec. D for a discussion of how the density depletion at a given optical potential varies with changing $a_\text{BF}$). Meanwhile, the Li cloud adiabatically follows the changing magnetic field due to the much faster time scale given by the Fermi energy as $h/E_\text{F} \approx 0.15$~ms. The theory curves in Figs. 3 and 4 are evaluated based on the assumption of a constant initial Cs profile. Equation 4 further assumes the Thomas-Fermi approximation and provides the magenta curves plotted in Figs. 3 and 4. A full simulation of the dynamics based on a coupled hydrodynamic model is presented in Sec. H.

The boson-boson scattering length $a_\text{BB}$ slightly varies over the range of magnetic fields that are studied in this work (see Fig. \ref{fig:avsb}). This contributes an overall variation in the background bare boson sound speed value $c_0$, which is not included in the presented theoretical predictions and would be interpreted as a sound speed shift in the experimental data. The sound speed change due to intraspecies scattering length variation is most significant at small $|a_\text{BF}|$, where $a_\text{BF}$ varies more slowly with magnetic field. This effect is negligible for the  majority of our data, but is likely responsible for the small drop in observed sound speed and increase in damping at small positive values of $a_\text{BF}$ in Figs.~3a and 3b.

There are three sources of uncertainty in determining the scattering lengths at which we perform our experiments. Our magnetic field settles to within 0.3\% of its final value within 5 ms of the switch. On our largest switches of 2 G, this corresponds to a 1-$\sigma$ uncertainty on the field of 6 mG during the sound propagation. The resolution of our determination of the magnetic field using microwave spectroscopy is 4 mG. And finally, our prior measurement of the Feshbach resonance position has an uncertainty of 1 mG \cite{Johansen2017}. Figs.~2, 3, 4, and S8 include error bars on the scattering length corresponding to these three sources of uncertainty added in quadrature ($7.3$ mG).

To make the measurements shown in Figs.~3c and 3d, we linearly ramp the trap depth down to a target value then back up to its original value over 400 ms. The number of Cs atoms that escape the trap depends on the target value, allowing control over the density. We confirm that this procedure does not result in appreciable heating of the bosons or loss of the fermions.

\subsection*{B. In situ imaging and DMD potential projection}

\begin{figure}
\center
\includegraphics[trim=1.6in 6.7in 2.7in 0in,clip,width=0.48\textwidth]{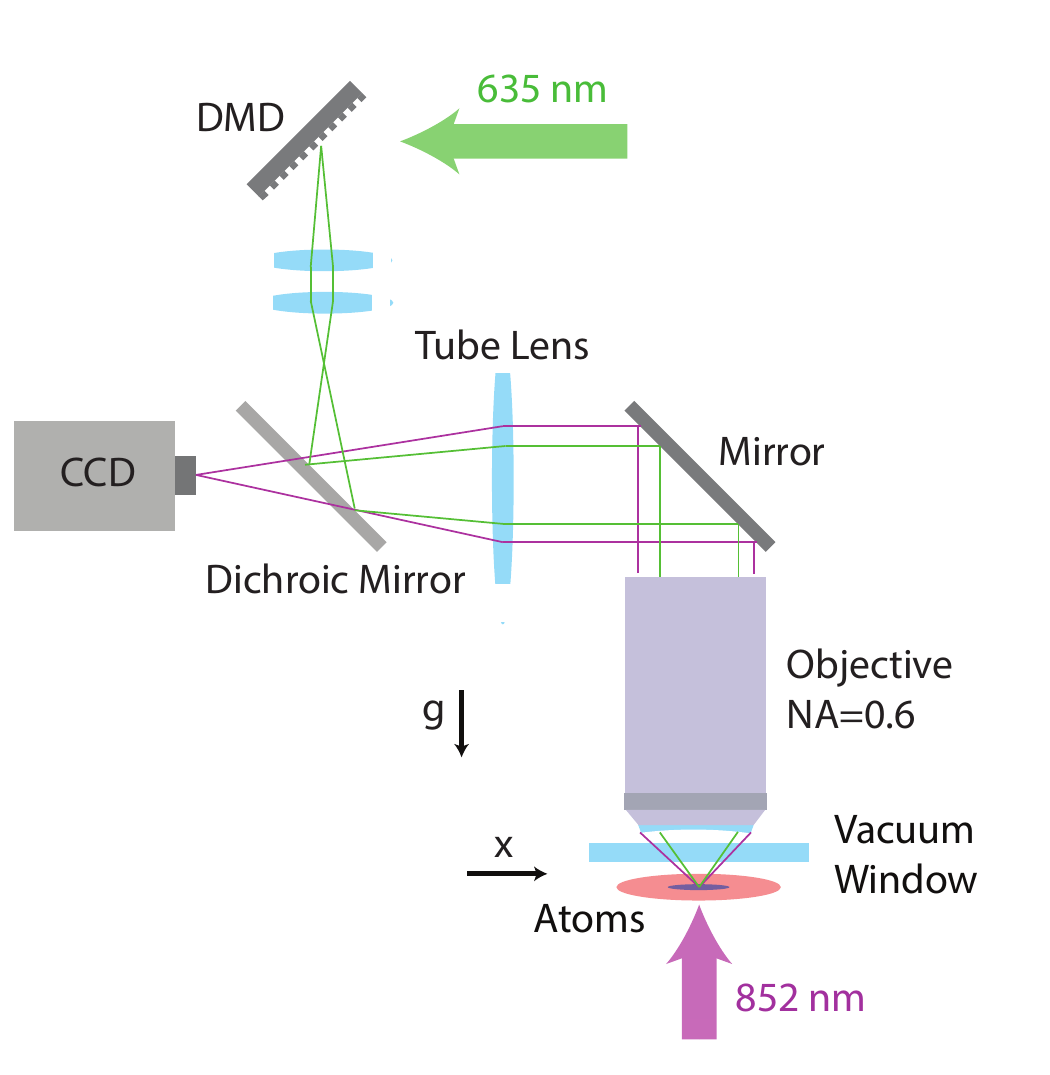}
\caption{\label{fig:microscope} Optical potential projection and Cs BEC imaging. A high resolution microscope objective with numerical aperture NA=0.6 is positioned close to the atomic sample. Imaging light near 852 nm (red arrow) is sent upwards through the objective, and is transmitted through a dichroic mirror and imaged onto a CCD. Light which is blue-detuned at 635 nm (green arrow) is reflected off a configurable digital micromirror device (DMD) before being reflected from the same dichroic and projected onto the atomic sample.}
\end{figure}

To obtain sufficient signal for absorption imaging of Cs, we first optically pump the atoms from $|F=3,m_F=3 \rangle$ to $|F=4,m_F=4 \rangle$, where we can take advantage of the cycling transition on the D2 line from $|F=4,m_F=4 \rangle\leftrightarrow |F'=5,m_F'=5\rangle$. The prime refers to excited states in the  6$^2$P$_{3/2}$ manifold. For simplicity, we label all states according to the quantum numbers of the low field Zeeman sublevels to which they are adiabatically connected.  We image the atoms by exposing them to 2\,\SI{}{\micro\second} of pumping light and 10\,\SI{}{\micro\second} of imaging light with an overlapping leading edge. Our imaging is performed at an intensity $I/I_{\text{sat}} \approx 6$, where $I_{\text{sat}}$ is the saturation intensity of the cycling transition.

We perform imaging using a custom microscope objective from Special Optics with numerical aperture NA=0.6. The microscope is designed for diffraction limited performance at the D2 line of both Cs (852 nm) and Li (671 nm). The image is then captured on a CCD camera (Andor iKon M 934), see Fig. \ref{fig:microscope}. To project the repulsive barrier onto the atoms, we reflect 635 nm light off a DMD (Texas Instruments DLP3000) and send it through the microscope using a dichroic mirror. With our imaging system, we resolve features down to 0.78(2)\,\SI{}{\micro\meter} for 852 nm imaging light, which is sufficient to resolve the $4$\,\SI{}{\micro\meter} wide density waves as they travel.

\begin{figure}[ht]\center
\includegraphics[trim=2.15in 3in 2.2in 3in,clip,width=0.48\textwidth]{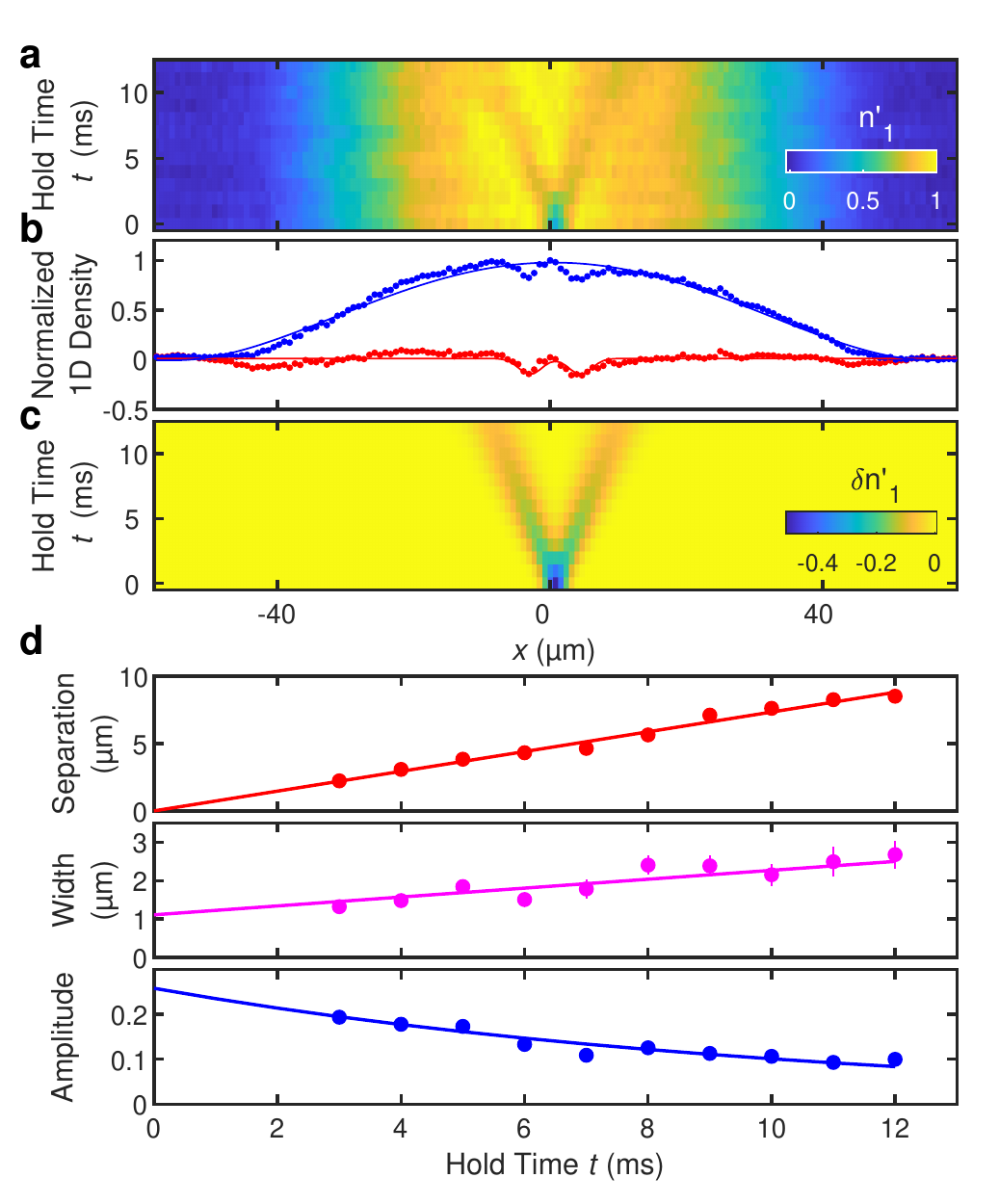}
\caption{\label{fig:data_analysis} Data analysis procedure for the extraction of sound speed and damping from the experimental data. Sample analysis is shown for interspecies scattering length $a_{\text{BF}}=-335\,a_0$. (a) Dynamics of normalized 1D profiles $n'_1$. (b) Sample profiles for $t=5$ ms. The measured 1D density $n'_1$ (blue circles) is fit to a bimodal fit function (blue  line, see text). The fit is subtracted off to produce the density wave profile $\delta n'_1$ (red circles). The  red line is obtained from the full 2D fit shown in (c). (d) Amplitudes, widths and separations of the density waves versus time extracted from $\delta n'_1$ compared for 2D fits (lines) and independent 1D Gaussian fits (circles). The lines are the separation $vt$, the width $\sigma_0+bt$ and the amplitude  $Ae^{-\Gamma t}$ from the 2D fit (see text). }
\end{figure}

\subsection*{C. Determination of density wave velocity and damping}

We extract velocities $v$ of the density waves  from images in two steps. For a given hold time $t$, we first integrate the images along the tight direction then normalize by the measured peak value to obtain $n'_1$ (see Fig.~\ref{fig:data_analysis}a). Then, we perform a bimodal fit of the density distribution according to the fit function

\[n_{\text{fit}}(x) = n_0\left[1-\left(\frac{x^2}{R_x^2}\right)\right]^2+ n_\text{th}e^{\frac{-x^2}{2\sigma^2}} + C,  \]

\noindent where $n_0$, $R_x$, $n_{\text{th}}$, $\sigma$ and $C$ are fit parameters capturing the peak 1D density of the condensate, the condensate Thomas-Fermi radius, the  thermal fraction 1D peak density, and an offset that accounts for possible detection noise. We subtract the fit from the data to obtain the density profile of the density waves $\delta n'_1 = n_1' - n_{\text{fit}}$ (see Fig.~\ref{fig:data_analysis}b).

We then perform a 2D fit to the evolution of the density wave profiles $\delta n_1'$ using the function 

\[\delta n_{\text{fit}}(x,t) = Ae^{-\Gamma t}\left[e^{\frac{-(x-x_0+vt)^2}{2(\sigma_0+bt)^2}}+e^{\frac{-(x-x_0-vt)^2}{2(\sigma_0+bt)^2}}\right] + C, \]

\noindent where $A$, $\Gamma$, $x_0$, $v$, $\sigma_0$, $b$ and $C$ are fit parameters representing the initial amplitude, the decay rate, the position of the initial depletion, the density wave velocity, the initial depletion width, the rate at which the depletion widens over time, and an offset that accounts for possible detection noise (see Fig.~\ref{fig:data_analysis}c). 

\begin{figure}[t]\center
\includegraphics[trim=2.2in 3.2in 2in 3.2in,clip,width=0.48\textwidth]{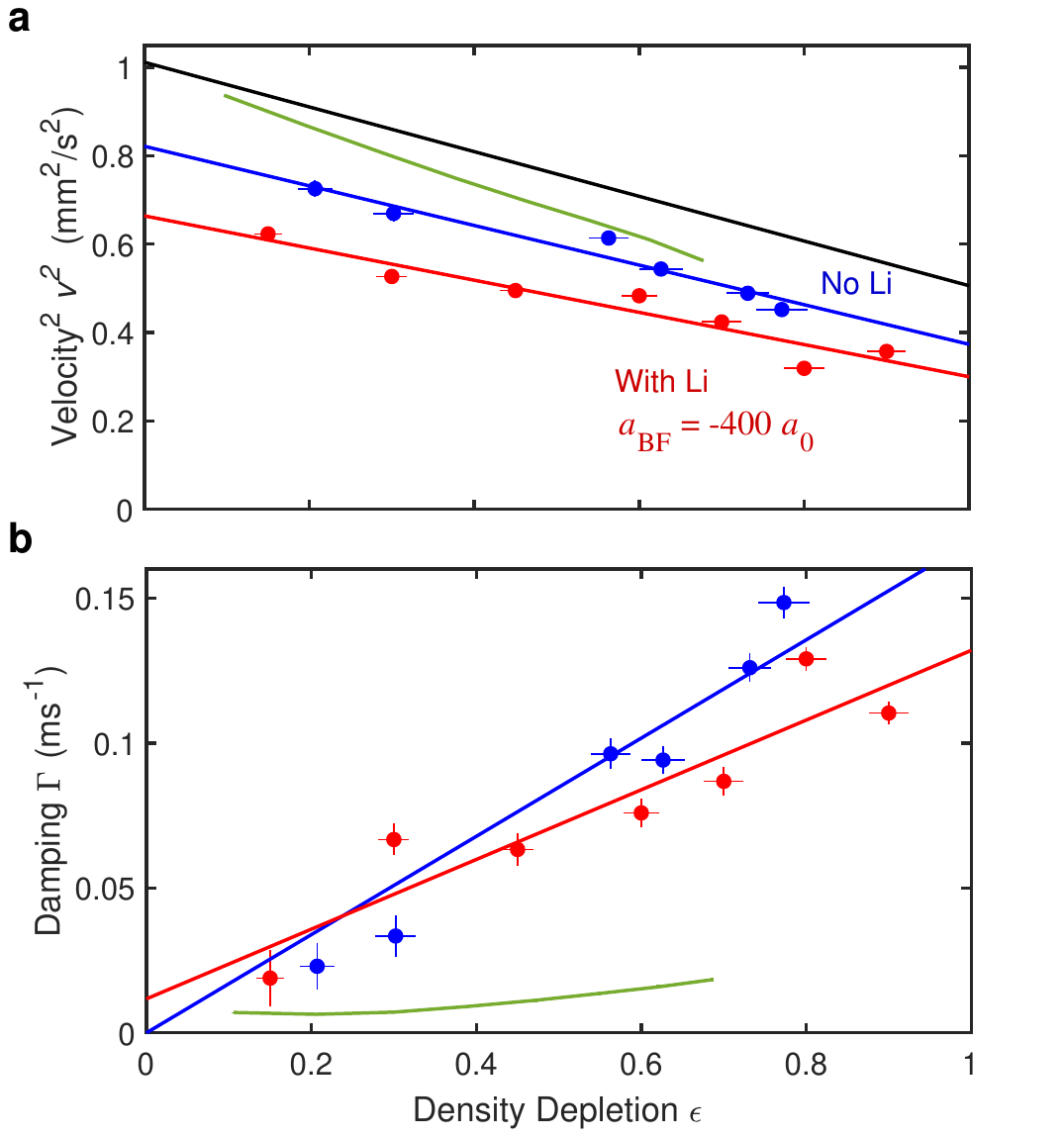}
\caption{\label{fig:depletion} Effect of initial depletion on density wave velocity and damping. (a) The measured  density wave velocity is shown for samples with no fermions present (blue circles) and in the presence of fermions with $a_\text{BF}=-400\,a_0$ (red circles).  In both cases, the dependence of the squared sound velocity is well captured by a linear fit (blue and red lines). The black line is the analytical result Eq. (\ref{depletionc}) and the green line is from the hydrodynamic simulations (see Section H), both evaluated in the absence of fermions. (b) Measured damping for the same data sets shown in the same colors. The blue and red lines are linear fits to the data. The green line is from the hydrodynamic simulation without fermions. All error bars are standard errors calculated from  fits to the experimental images. }
\end{figure}

The fit function $\delta n_{\text{fit}}$ assumes constant velocity motion of the depletions, an exponential decay of their amplitude, and a linear increase in their width. These constraints are chosen based off the observed behavior of the depletions when the density wave profile $\delta n_1'$ for each hold time is fit independently by a pair of Gaussian functions. The extracted amplitudes, widths, and separations of the waves from 2D fits and independent 1D fits are compared in Fig. \ref{fig:data_analysis}d, for hold times after the peaks have separated enough to yield reliable results for both methods. 

While both methods yield compatible results for the extracted velocities, we find that the full 2D fits are more robust against noise in the data. Additionally, performing the 2D fits allows us to extract information from early times before the two peaks have become separated, permitting  measurement of the damping rate $\Gamma$ without additional assumptions.

The background subtraction process is imperfect, due to large length scale variation of the BEC density profile. This corresponds to some uncertainty in which parts of the density profile are the density wave and which parts are the background. We attribute an uncertainty of $5\%$ to this systematic, which is estimated by comparing extracted velocities for different viable background subtractions. This is the largest estimated systematic uncertainty in our analysis.

We determine whether the sound propagation is stable based on the evolution of the density profiles. We examine the normalized 1D density profiles at each time step individually. If two separated depletions can be observed in most of these profiles at later times, we report a sound velocity based on the fit. If most of the profiles do not show two depletions, we consider the sound mode unstable. All data sets fall into the above two categories. For a comparison of these two cases see Fig.~\ref{fig:supp_efimov}.

\begin{figure}[t]\center
\includegraphics[trim=2.2in 3.85in 2.5in 4in,clip,width=0.48\textwidth]{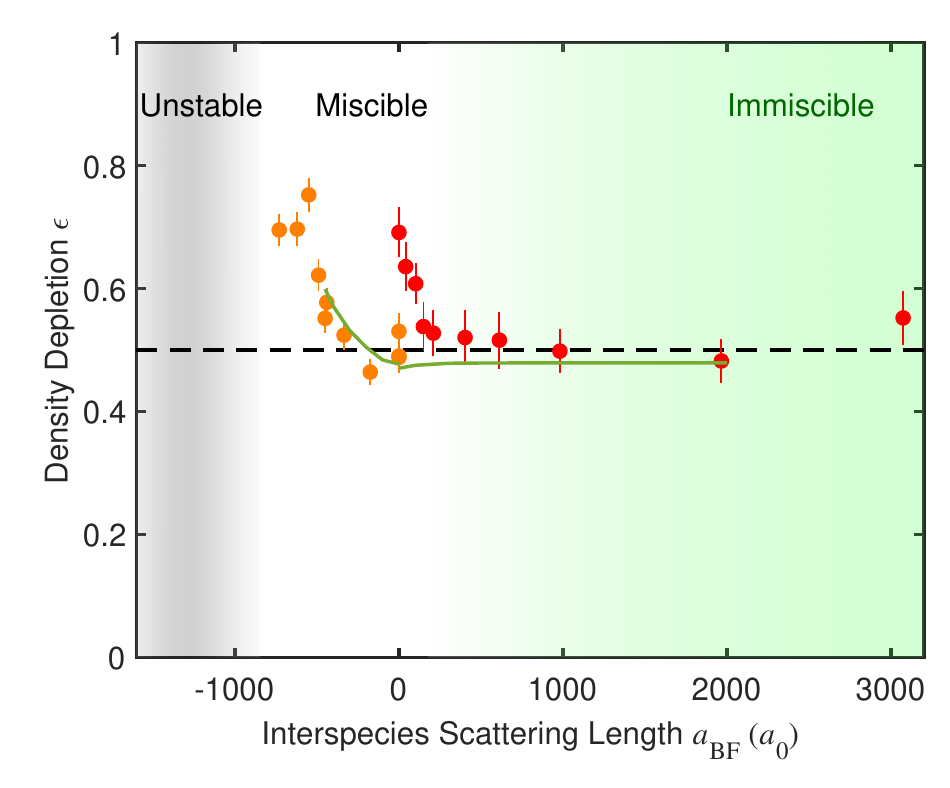}
\caption{\label{fig:depletion_abf}  Change in initial depletion  due to interspecies interactions. Orange and red circles are data from the attractive and repulsive side, respectively. The green solid line is simulated from the hydrodynamic model (see Section H). The black dashed line indicates the initial experimentally set depletion $\epsilon= 0.5$ for interspecies scattering length $a_{\text{BF}}=-150 \,a_0$. The region shaded in grey indicates where we observe unstable sound propagation, and the region in green indicates phase separation of the components. Error bars are standard errors calculated from fits to averaged density profiles.}
\end{figure}

\subsection*{D. Dependence of density wave dynamics on depletion}

In our elongated geometry, density waves propagating along the long axis of the condensate can be described as waves in the 1D density $n_1$ that travel with a velocity $v_0$ \cite{Kavoulakis1998}, given by

\[v_0^2 = \frac{\bar{n}_\text{B}g_\text{BB}}{m_{\text{B}}} = c_0^2\frac{\bar{n}_\text{B}}{n_\text{B}}, \]
where $\bar{n}_\text{B} = n_1/A$ is the mean 3D density over the transverse cross-section $A$ and $n_\text{B}$ is the 3D density evaluated along the symmetry axis. For harmonic transverse confinement, the Thomas-Fermi approximation gives $\bar{n}_\text{B} = n_\text{B}/2$ and thus $v_0 = c_0/\sqrt{2}$. 

 For a density wave with significant density depletion $\delta n_1$, the propagation speed is reduced due to the lower mean density.  Assuming the cross-section $A$ is constant during the propagation, the density wave velocity $v$ is given by

 \[v  \approx \frac{c_0}{\sqrt{2}} \sqrt{1-\frac{\delta n_1}{n_1}} \equiv  \frac{c_0}{\sqrt{2}}\sqrt{ 1 - \frac{\epsilon}{2}}, \]

\noindent where $\epsilon \equiv 2\delta n_1/n_1$ is the fractional depletion of the 1D density induced by the optical barrier. The factor of $2$ accounts for the splitting of the initial density depletion into two equal amplitude density waves propagating in opposite directions.

 We compare the measured density wave velocities $v$  to this prediction  by varying the optical power in the potential barrier, see Fig. \ref{fig:depletion}a. The depletion $\epsilon$ is extracted from a single Gaussian fit to the initial perturbation density profile $\delta n_1'$ at hold time $t=0$. In the absence of fermions, we find fair agreement with Eq. (\ref{depletionc}). A linear fit to the squared velocity $v^2 = \nu_0^2 + m\epsilon$ gives $\nu_0 =0.91(2)$ mm/s, and thus $c_0 = 1.28(3)$ mm/s. This value is consistent within 10\% of our estimate. From the fit, we determine the slope to be $m=-0.5(1)\nu_0^2$ consistent with the prediction $-0.5\,\nu_0^2$.
 
The same experiment in the presence of fermions at scattering length $a_\text{BF} = -400\,a_0$ with similar particle number yields an overall reduction of the density wave velocity. Using the same fit function, we obtain $\nu_0 = 0.81(2) $ mm/s and the slope $m=-0.5(1)\nu_0^2$, consistent with Eq. (5).

\begin{figure*}\center
\includegraphics[width=\textwidth]{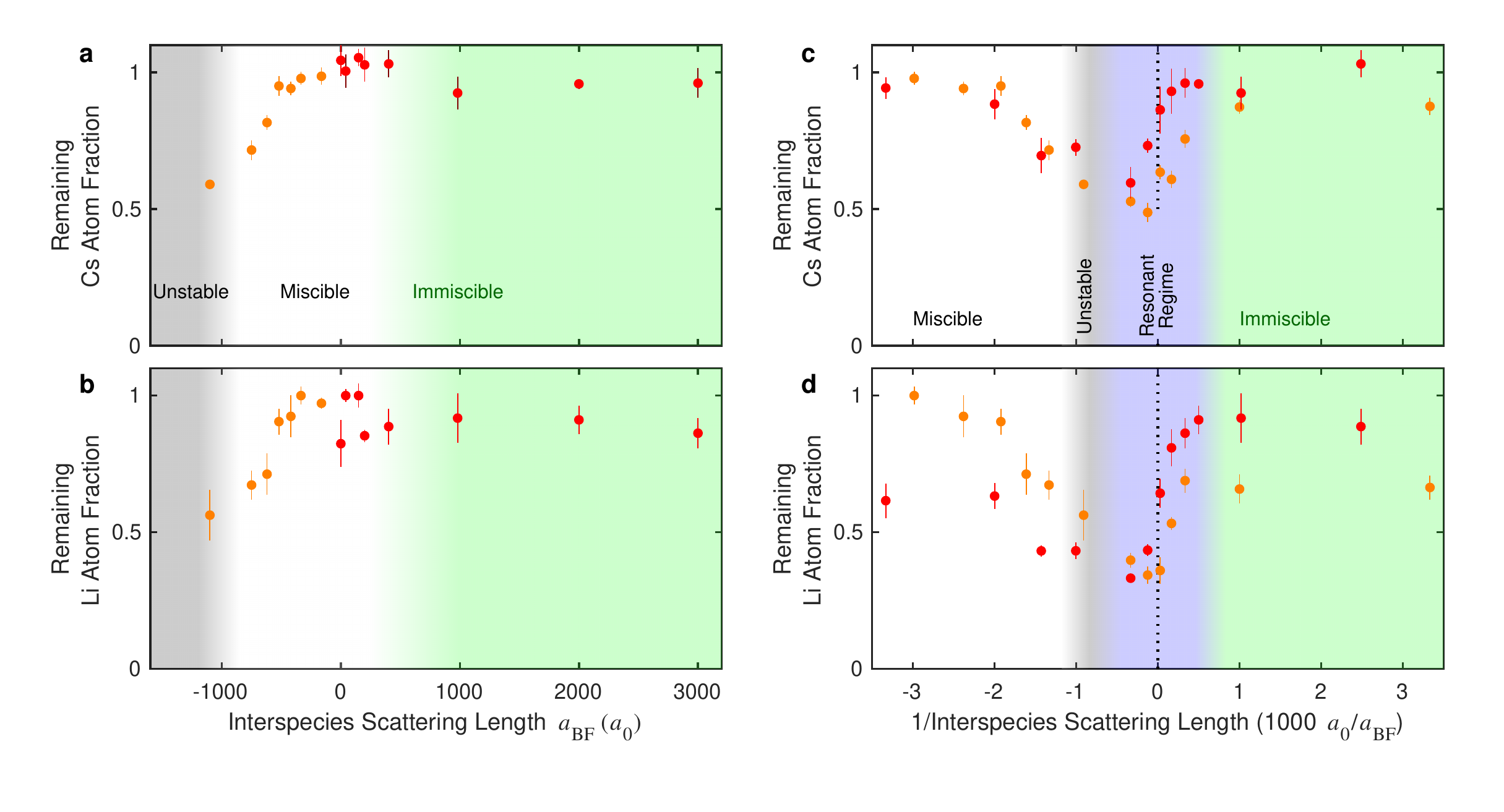}
\caption{\label{fig:loss} Loss of Cs and Li atoms across the Feshbach resonance. Experimental data showing the fraction of Cs atoms (panel (a) and (c)) and Li atoms (panel (b) and (d)) remaining after a hold time of $t=6$ ms. The orange and red circles indicate experiments performed by preparing on the attractive and repulsive side of the Feshbach resonance, respectively. The Cs atom data is taken after a 4 ms time of flight, and the Li data is taken \textit{in situ}.  The region shaded in grey indicates the region where no stable sound propagation is observed. The region shaded in green indicates the phase separated region. The region indicated in blue indicates the resonant regime. The dotted line in panels (c) and (d) indicate the position of the Feshbach resonance. Error bars are the standard error.  }
\end{figure*}

The initial depletion $\epsilon$ also has a significant effect on the damping rate, as shown in Fig. \ref{fig:depletion}b. This effect is much weaker in our simulation, which suggests that it does not come from the nonlinearity present in that model. To characterize the damping quantitatively, we perform a linear fit to each data set while constraining the y-intercept to be positive. This gives a slope $0.17(2)$ ms$^{-1}$ without Li and  $0.12(2)$ ms$^{-1}$ with Li at scattering length $a_\text{BF}=-400 a_0$.

\begin{figure*}
    \center
    \includegraphics[trim=0in 1.1in 0in 0.2in,clip,width=\textwidth]{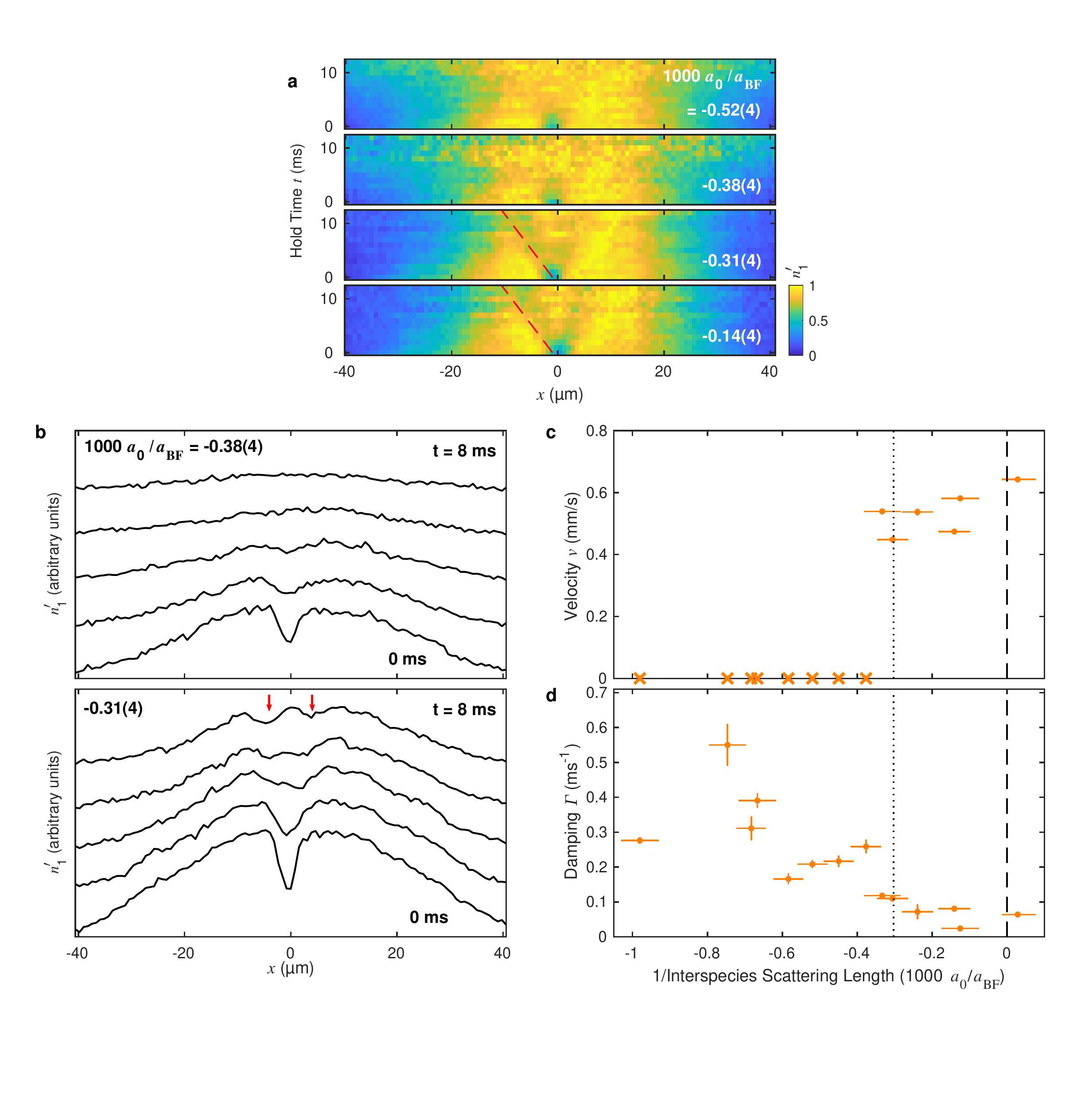}
    \caption{Reemergence of sound propagation. (a) Normalized 1D densities illustrating the revival of sound propagation between $1000\, a_0/a_\text{BF} = -0.38$ and $-0.31$ based on the same experimental procedure as in Figs.~\ref{fig:cvsabf} and \ref{fig:cres}. For all data, the system is prepared on the attractive side of the Feshbach resonance and switched to the target magnetic field. Red dashed lines are guides to the eye. (b) Normalized 1D densities at $t = 0$, 2, 4, 6, and 8 ms after the onset of density wave dynamics. In the data at $1000\, a_0/a_\text{BF} = -0.31$, two clearly separated density depletions are visible at later times (red arrows), whereas no such structure is visible in the data at $1000\, a_0/a_\text{BF} = -0.38$.
    (c) Density wave velocity near the Feshbach resonance. The crosses indicate samples with no stable sound propagation. (d) Damping from the same data set. The vertical error bars in panels (c)-(d) are standard errors from fits to averaged density profiles. The horizontal error bars represent 1-$\sigma$ uncertainties of the scattering length (see Section A). The vertical dashed line shows the position of the Feshbach resonance. The vertical dotted line shows the position of the Efimov resonance reported in Ref.~\cite{Johansen2017}.}
    \label{fig:supp_efimov}
\end{figure*}

For the data in the main text figures, we prepare our gas with an initial density depletion of $\epsilon = 0.5\pm .05$ (see Fig.~\ref{fig:sampledata}) near the initial sympathetic cooling field, prior to our magnetic field ramp at time $t=-5$ ms. However, due to the change in interspecies interactions during the ramp, the level of depletion when the optical barrier is switched off can vary as a function of the target interspecies scattering length. This variation for the data in Figs. 3a and 3b  is shown in Fig.~\ref{fig:depletion_abf}. 

The increase in the depletion $\epsilon$ at negative $a_\text{BF}$ is likely due to interspecies interactions, but the increase at small positive $a_\text{BF}$ is likely due to the reduction of the intra species scattering length $a_\text{BB}$ for those data points (see Fig.~\ref{fig:avsb}). Comparing the results in Fig.~\ref{fig:depletion} to those in Fig.~\ref{fig:depletion_abf} suggests that this depletion dependence can contribute about 10\% additional reduction in the depletion velocity $v$ and an additional $0.05$ ms$^{-1}$ to the damping rate near the unstable region where the change in density depletion is greatest.

In Figs. 3b and 4b we compare the experimental data to a perturbation prediction \cite{Viverit2002}

\[\Gamma = \text{Im}[\omega] =  kc_0\frac{\sqrt{\pi}(1+w)^2}{4w^2}\frac{a_\text{BF}^2}{a_\text{BB}^2}\sqrt{n_\text{B}a_\text{BB}^3},  \]

\noindent where $w = \frac{m_\text{B}}{m_\text{F}}$ is the mass ratio between the species. To provide comparison to the data, we evaluate the perturbation prediction for a phonon momentum $k=\frac{2\pi}{\delta}$, where $\delta$ is the width of the density waves. Our measurements do not distinguish the origin of the damping, but we note that the maximum contribution from the measured depletion dependence is about 0.15 ms$^{-1}$, which is smaller than the largest measured values shown in Figs. 3 and 4.

\subsection*{E. Atom loss}

The atom loss rate due to three-body recombination can change both due to changes in the three-body loss rate coefficient and the overlap of the two species. The high atomic densities in the \textit{in situ } images lead to complications in the direct determination of the atom number. 

We therefore perform a complimentary experiment where we control the interspecies interactions identically as for the data in the main text, but do not excite sound waves using the optical barrier. We count the atom number after a hold time of $t=6$ ms, imaging the Cs atoms after a 4 ms time of flight expansion and the Li atoms \textit{in situ}. The results are presented in Fig.~\ref{fig:loss}. The peak of the loss in each case appears close to the pole of the Feshbach resonance, rather than in the region  highlighted in grey where we do not see stable sound propagation.

We note that the measured velocities at resonance are slightly lower for samples prepared on the negative side compared to the positive side (see Fig.~\ref{fig:cres}). We attribute this to the stronger particle loss on the attractive side of the resonance.




\subsection*{F. Re-emergence of sound propagation}

We perform additional sound speed measurements to determine the precise conditions for the reemergence of sound propagation. Based on similar experimental conditions as in Fig. 4 and magnetic field control at higher resolution, we observe a stark reemergence of the sound propagation between $a_\text{BF} = 2600\, a_0$ and $3200\, a_0$ ($1000\, a_0/a_\text{BF} = -0.38$ and $-0.31$) based on the criterion described in Sec. C, see Fig. S8. The transition is very close to the position of the Efimov resonance at $a_\text{BF} = -3300\, a_0$ ($1000\, a_0/a_\text{BF} = -0.30$) \cite{Johansen2017}. Future investigation is needed to determine the role of the Efimov resonance in the reemergence of the sound propagation.

\subsection*{G. Phonon-fermion coupling}

In this section, we briefly summarize a portion of the discussion of Ref. \cite{Viverit2002} to obtain the phonon-fermion coupling in the Bose-Fermi mixture. We start from the Hamiltonian for the uniform mixture in the Bogoliubov approximation

\[H =  \sum_k\epsilon_k^\text{F}c_k^{\dagger}c_k + E_\text{B} + \sum_k\hbar\omega_k\alpha^{\dagger}_k\alpha_k + g_\text{BF}\int d\vec{r}\,\, n_\text{B}n_\text{F}, \]

\noindent where $\epsilon_k^\text{F} = \frac{\hbar^2k^2}{2m_\text{F}}$ is the kinetic energy of a fermion, $E_\text{B}$ is the ground state energy of the bosons, $n_\text{F}$ is the average fermion density, $n_\text{B}$ is the average boson density, and  $\hbar\omega_k = \sqrt{\left( \epsilon_k^\text{B}\right)^2 + 2g_\text{BB}n_{\text{B}}\epsilon_k^\text{B} }$ is the Bogliubov dispersion, where $\epsilon_k^\text{B} = \frac{\hbar^2k^2}{2m_\text{B}}$ is the kinetic energy of a boson. The first term is the total kinetic energy of the Fermi gas, the next two terms are the energy of the Bose gas, and the last is the total interaction energy between the two components.

The annihilation and creation operators  $\alpha_k$ and $\alpha^{\dagger}_k$ for the phonons are related to the corresponding operators for the bosonic atoms $a_k$ and $a^{\dagger}_k$ through the Bogoliubov transformation
\[a_k = u_k\alpha_k + v_k\alpha^{\dagger}_{-k} \]
\[a^{\dagger}_k = u_k\alpha^{\dagger}_k + v_k\alpha_{-k},\]
with coefficients defined by
\[ u_k^2 = \frac{1}{2}\left( \frac{\epsilon_k^\text{B} + g_\text{BB}n_\text{B}}{\hbar\omega_k} + 1 \right) \]
\[ v_k^2 = \frac{1}{2}\left( \frac{\epsilon_k^\text{B} + g_\text{BB}n_\text{B}}{\hbar\omega_k} - 1 \right). \]
 The interaction term is rewritten by defining density fluctuation operators at momentum $k$

\[\rho_k^\text{B} = \frac{1}{\sqrt{V}} \int d\vec{r}\, e^{i\vec{k}\cdot\vec{r}} [n_\text{B}(\vec{r}) - n_\text{B}] = \frac{1}{\sqrt{V}}\sum_q a^{\dagger}_qa_{q+k}  \]
\[\rho_k^\text{F} = \frac{1}{\sqrt{V}} \int d\vec{r}\, e^{i\vec{k}\cdot\vec{r}}[ n_\text{F}(\vec{r}) - n_\text{F} ] = \frac{1}{\sqrt{V}}\sum_q c^{\dagger}_qc_{q+k}. \]

\noindent where $n_\text{B}(\vec{r})$ and $n_\text{F}(\vec{r})$ are the local boson and fermion densities, respectively. The ground state energy of the bosons is $E_\text{B} = g_{\text{BB}}n_\text{B}N_\text{B}/2$, where $N_B$ is boson number.  The interaction term is now:
\[g_\text{BF}\int d^3r n_\text{B}n_\text{F}  = g_{\text{BF}}N_\text{B}n_\text{F} + g_\text{BF}\sum_k\rho_k^\text{B}\rho_{-k}^\text{F}.\]

\noindent Using the Bogoliubov approximation, the boson density fluctuation operator is given by
\[\rho_k^\text{B} \approx \sqrt{n_\text{B}\hbar k^2/2m_\text{B}\omega_k}(\alpha_k + \alpha_{-k}^{\dagger}).\]

\noindent Substituting this into the interaction term, we arrive at the Hamiltonian in the form
\begin{align*}
H = &E_0 + \sum_k\epsilon_k^\text{F}c_k^{\dagger}c_k +\sum_k\hbar\omega_k\alpha_k^{\dagger}\alpha_k   \\ &+ g_\text{BF}\sqrt{n_\text{B}}\sum_{k,q}\sqrt{\hbar k^2/2m_\text{B}\omega_k}(\alpha_k+\alpha^{\dagger}_{-k})c^{\dagger}_qc_{q-k},
\end{align*}

\noindent which is Eq. (1) referred to in the main text with the explicit form of the coupling $g_k$ and overall energy offset $E_0 = E_\text{B} + g_{\text{BF}}N_{\text{B}}n_{\text{F0}}$ included.

\subsection*{H. Coupled hydrodynamic model and numerical simulation}

\begin{figure}\center
\includegraphics[trim=2.2in 3.5inn 2.3in 3.5in,clip,width=0.48\textwidth]{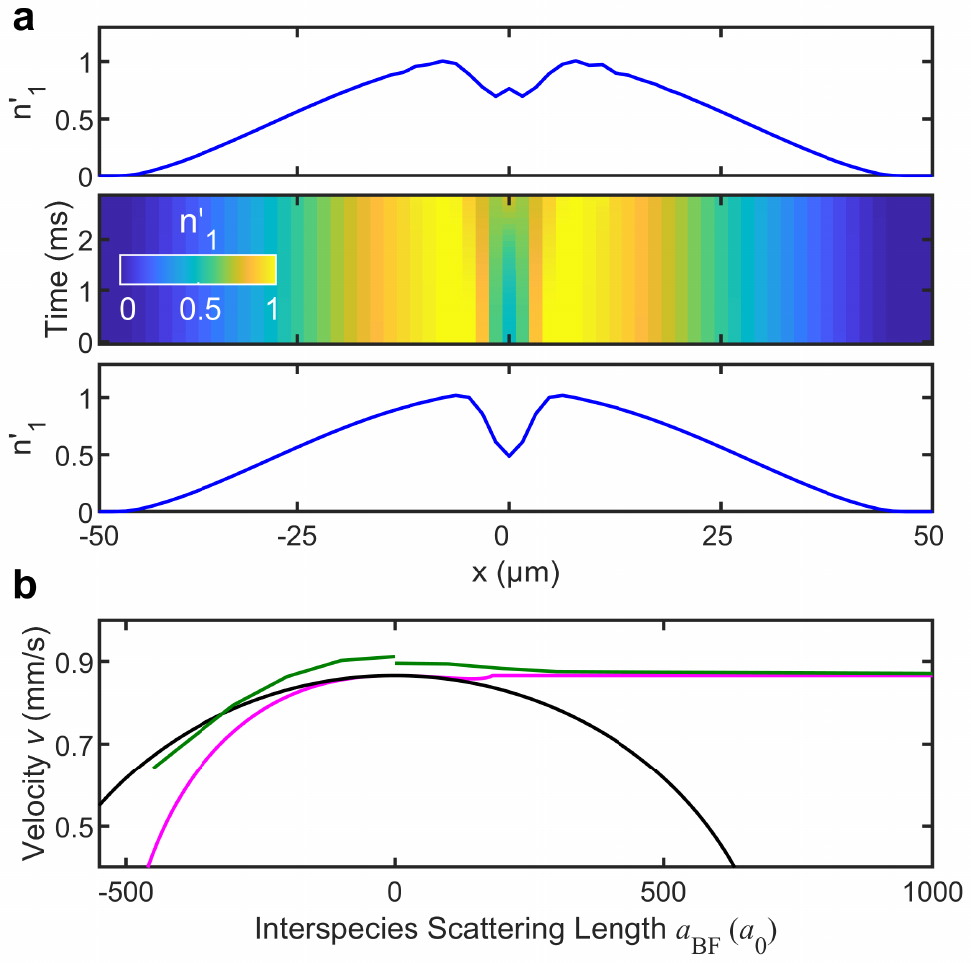}
\caption{\label{fig:sim} Coupled hydrodynamic simulations and comparison of all models. (a) Sample normalized 1D density profiles $n'_1$ generated from the coupled hydrodynamic simulations for $a_\text{BF} = -300\, a_0$. The lower and upper profiles are for $t=0$ ms and $t=2.8$ ms respectively. (b) Density wave velocities fit from 3D simulations of the experimental dynamics using the coupled hydrodynamic model are shown as the green line. The discontinuity at $a_\text{BF}=0$ is due to different preparation of the sample for $a_{\text{BF}}>0$ and $a_{\text{BF}}<0$ in the numerical simulation (see text). The result is compared with the perturbation  (black line) , see Eq. (2), and mean-field prediction (magenta line), see Eq. (\ref{mfeq}).}
\end{figure}

A coupled hydrodynamic model that corresponds to a typical Gross-Pitaevskii equation for bosons and a hydrodynamic treatment for fermions can be used to approximate the long wavelength dynamics of the system. This type of treatment has been studied in detail theoretically \cite{Rakshit2019,Karpiuk2020}. The system is described by a pair of equations:

\begin{align*}
	i\hbar\frac{\partial\psi_\text{B}}{\partial t}=\bigg(-\frac{\hbar^2}{2m_\text{B}}\nabla^2&+g_\text{BB}|\psi_\text{B}|^{2}+g_\text{BF}|\psi_\text{F}|^2 \bigg)\psi_\text{B}\\
	i\hbar\frac{\partial\psi_\text{F}}{\partial t}=\bigg(-\frac{\hbar^2}{2m_\text{F}}\nabla^2&+\frac{\hbar^2\xi'}{2m_\text{F}}\frac{\nabla^2|\psi_\text{F}|}{|\psi_\text{F}|}
	+\frac{5}{3}\kappa_{\text{F}}|\psi_\text{F}|^{4/3}\\+g_\text{BF}|\psi_\text{B}|^2 \bigg)\psi_\text{F},
\end{align*}

\noindent where $\psi_\text{B}$ is the condensate wavefunction, $\xi' = 8/9$ is a factor arising from the Von Weiszacker gradient correction \cite{Kirznits1957}, $\kappa_{\text{F}} = (3/10)(6\pi^2)^{2/3}\frac{\hbar^2}{m_\text{F}}$ and $\psi_\text{F} = \sqrt{n_\text{F}}e^{i\phi}$ is the fermion pseudo-wavefunction where the local velocity of fermions is $-\frac{\hbar}{m}\nabla \phi$. We evaluate the model using the split operator method \cite{Gawryluk2018}. In Fig.~\ref{fig:sim}a, we show an example of 1D densities $n'_1$ vs hold time $t$  numerically calculated using this model. The dynamics are similar to those seen in the experiment (see Fig.~\ref{fig:sampledata}).

We perform two sets of numerical simulations which capture the different experimental procedures on the negative and positive side of resonance. For these simulations,  the intraspecies scattering length is held fixed at $a_{\text{BB}}=270 a_0$, but the dynamics of the interspecies scattering length $a_{\text{BF}}$  during the experimental magnetic field ramp are included. The evolution of the system is simulated for the duration of the 5 ms magnetic field ramp then 2.8 ms of evolution after the optical quench. A fit is performed to the simulated density wave dynamics to obtain the density wave velocity. There is an estimated systematic error of $5$\% in the obtained density wave velocity that arises from the choice of simulation duration.  The discontinuity at $a_{\text{BF}}=0$ is due to small changes in the boson density resulting from the different initial values of the scattering length $a_{\text{BF}}$ at $t=-5$ ms (see Section A).

In Fig.~\ref{fig:sim}b we compare the predicted sound velocities from the hydrodynamic simulations to Eq. (\ref{rkkyc}) and (4)  discussed in the main text.  We see fair agreement between our analytical models and the hydrodynamic simulations. We note that for our system parameters, our simulations suggest that there is no stable ground state for $a_\text{BF} < -450 \, a_0$, which we identify with instability towards collapse.

\subsection*{I. Thomas-Fermi approximation}

In this section, we provide a more detailed explanation Eq. (3) and (4) in the main text. We use a Thomas-Fermi approximation for both species, which allows us to ignore all of the spatial gradient terms in the hydrodynamic model. We write the chemical potential of each species

\[\mu = g_\text{BB}n_\text{B} +  V_\text{B} + g_\text{BF}n_\text{F} \]
\[E_\text{F} = \frac{\hbar^2}{2m_\text{F}}(6\pi^2n_\text{F})^{2/3} + g_\text{BF}n_\text{B} + V_\text{F},\]

\noindent where $V_\text{B}$ and $V_\text{F}$ are the external potentials felt by the bosons and fermions, respectively. If the number of fermions pulled into or pushed out of the BEC by the interspecies interaction is small compared to the total fermion number, the fermion chemical potential $E_\text{F}$ is close to its bare value  $E_{\text{F0}}$, so $ E_{\text{F}} \approx E_\text{F0}$, where $E_\text{F0} = \frac{\hbar^2}{2m_\text{F}}(6\pi^2n_\text{F0})^{2/3}$. This approximation is justified in our experiment because only about 1$\%$ of the fermions are overlapped with the BEC for our system parameters.

 We consider the local chemical potential of the bosons, and define $\mu_{\text{TF}} = \mu - V_B$. Combining the expressions for the bosonic and fermionic chemical potentials yields

\[\mu_{\text{TF}} =  g_\text{BB}n_\text{B} + g_\text{BF}n_\text{F0}\left(1-\frac{g_\text{BF}n_\text{B}}{E_\text{F0}}\right)^{3/2}, \]
when $g_{\text{BF}}n_\text{B}/E_{\text{F0}} < 1$, as in Eq. (3) in the main text. This expression provides the relationship between the local boson density $n_\text{B}$ and the local fermion density in the absence of the condensate $n_{\text{F0}}$ \cite{Huang2020}.

To provide the mean-field prediction for the effective two- and three-body coupling constants $g_2$ and $g_3$ in Fig. 3d, we expand the chemical potential $\mu_{\text{TF}}$ to second order in $n_{\text{B}}$, which gives

\[\mu_{\text{TF}} \approx g_2n_\text{B} + g_3n_{\text{B}}^2 + ..., \]

\noindent  where $g_2=g_{\text{BB}} - \frac{3g_{\text{BF}}^2n_{\text{F0}}}{2E_{\text{F0}}} $ and $g_3 = \frac{3g_{\text{BF}}^3n_{\text{F0}}}{8E_{\text{F0}}^2}$. These coupling constants can be expressed as an effective two-body scattering length $a_{\text{eff}} = a_{\text{BB}} - \frac{m_\text{B}m_{\text{F}}}{m_r^2}\frac{k_\text{F}}{2\pi}a_{\text{BF}}^2$ and an effective scattering hypervolume $\nu_\text{eff}  = 2\pi \frac{m_\text{B}m_\text{F}^2}{m_r^3}k_F^{-1}a_{\text{BF}}^3$.

We evaluate the density wave velocity from the compressibility and Eq. (5) of the main text. We obtain the final expression for the density wave velocity along the symmetry axis of an elongated Bose-Fermi mixture as

 \begin{figure}\center
\includegraphics[trim=2.3in 3.85in 2.4in 4in,clip,width=0.48\textwidth]{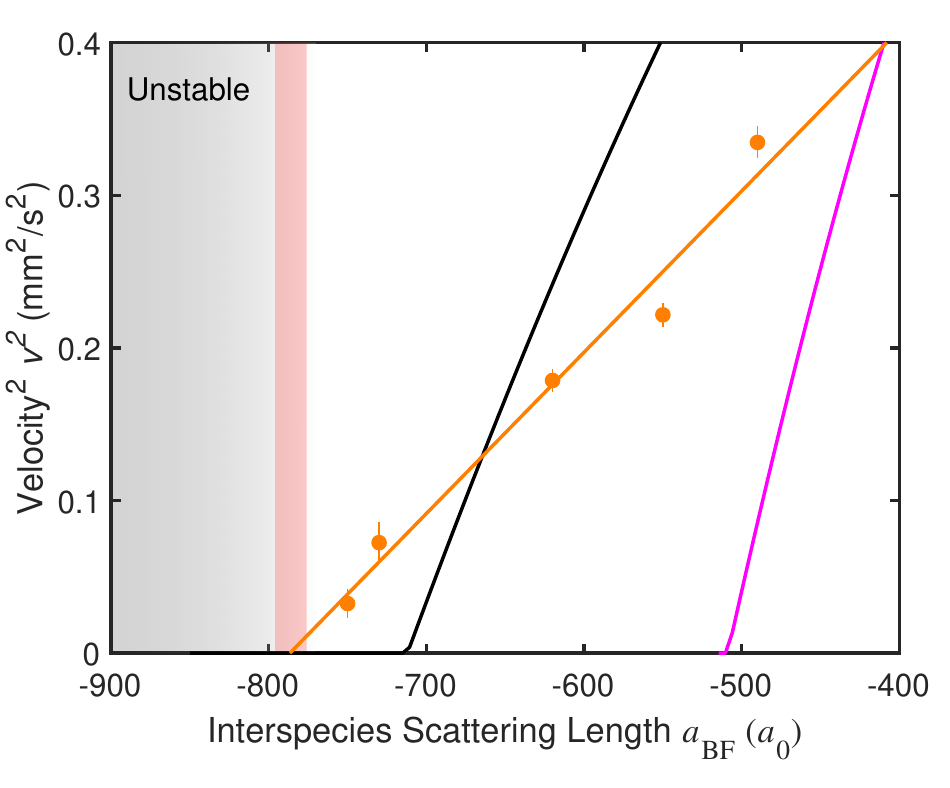}
\caption{\label{fig:crita} Density wave velocity near instability transition. The measured density wave velocities on the attractive side near the instability transition (orange circles) are fit to $v=\alpha\sqrt{a_{\text{BF}}-a_\text{c}}$ (orange line) to determine the critical scattering length. The red shaded area represents the standard error from the fit. For comparision, the perturbation (black line) and mean-field (magenta line) predictions are also shown. The grey area indicates the region where no stable sound propagation is observed. }

\end{figure}

\begin{equation}
\label{mfeq}
 v = \frac{c_0}{\sqrt{2}}\sqrt{1-\frac{\epsilon}{2}}\sqrt{1 - \frac{3}{2}\frac{g_\text{BF}^2}{g_\text{BB}}\frac{n_\text{F0}}{E_\text{F0}}\sqrt{1 - \frac{g_\text{BF}n_\text{B}}{E_\text{F0}}}}.
\end{equation}

 \noindent This expression is used to generate the mean-field predictions (magenta lines) in Figs. 3, 4, \ref{fig:sim}, and \ref{fig:crita}. To do so, we  approximate the fermion density $n_\text{F0}$ and the boson density $n_\text{B}$ as the bare peak densities of each species. We use the peak densities because the densities do not change significantly over the axial distance that the density waves propagate. We additionally make the assumption that the peak boson density is unchanged as we change the interspecies scattering length.
 
 In the above derivation, we assume that there is no significant radial motion of the condensate. In our experiment, we do not see clear signs of such dynamics. Furthermore, the sound speed evaluated from 3D hydrodynamic simulations  (see Fig. \ref{fig:sim}) is in fair agreement with this simplified mean-field model.

We include shaded regions on our theory curves corresponding to the range of the predictions due to experimental density
variations. Our Cs atom number varies at the $10\%$ level shot-to-shot, which results in a $2\%$ error on the sound speed. The local Cs density varies by $5\%$ over the course of the sound propagation, which results in a $2.5\%$ variation in the sound speed. The shaded regions correspond to these two uncertainties added in quadrature ($3.2\%$).

\subsection*{J. Determination of the critical scattering length}

As the interspecies attraction increases, both perturbation theory, see Eq. (2), and mean-field theory, see Eq. (7), predict the softening of the sound mode as $v \propto \sqrt{a_{\text{BF}}-a_\text{c}}$ , where $a_\text{c}$ is the critical scattering length. This dependence well captures the behavior of the data near the transition to the region of unstable sound propagation (see Fig.~\ref{fig:crita}).  We perform a fit to the 5 lowest density wave velocitiy measurements using the fit function $v = \alpha\sqrt{a_{\text{BF}} - a_{\text{c}}}$, where $\alpha$ and $a_\text{c}$ are fit parameters representing the velocity scale and the critical scattering length. The fit yields the critical scattering length $a_{\text{c}}=-790(10)\,a_0$ and the coefficent $\alpha = 32(1)$ \SI{}{\micro\meter}/s $\times\, a_0^{-1/2}$. The $10\,a_0$ error on $a_c$ is the statistical uncertainty from the fit. The 1-$\sigma$ systematic uncertainty on $a_c$ is $30\,a_0$.

\end{document}